\documentclass[12pt,preprint2]{aastex}
\usepackage{graphicx}
\shorttitle{Density intermittency in the fast solar wind}
\shortauthors{Bruno et al.}
\begin{document}
\title{Radial evolution of intermittency of density\\
fluctuations in the fast solar wind}
\author{R. Bruno\altaffilmark{1}, D. Telloni\altaffilmark{2}, L. Primavera\altaffilmark{3}, E. Pietropaolo\altaffilmark{4}, R. D'Amicis\altaffilmark{1}, L. Sorriso-Valvo\altaffilmark{3,5,6}, V. Carbone\altaffilmark{3}, F. Malara\altaffilmark{3} and P. Veltri\altaffilmark{3}}
\altaffiltext{1}{National Institute for Astrophysics, Institute for Space Astrophysics and Planetology, Via del Fosso del Cavaliere 100, 00133 Roma, Italy}
\altaffiltext{2}{National Institute for Astrophysics, Astrophysical Observatory of Torino, Via Osservatorio 20, 10025 Pino Torinese, Italy}
\altaffiltext{3}{University of Calabria, Department of Physics, Via Ponte P. Bucci Cubo 31/C, 87036 Rende, Italy}
\altaffiltext{4}{University of L'Aquila, Department of Physics, Via Vetoio Localit$\grave{\textrm{a}}$ Coppito, 67100 L'Aquila, Italy}
\altaffiltext{5}{IPCF-CNR, UOS of Cosenza, Via Ponte P. Bucci Cubo 33/B, 87036 Rende, Italy}
\altaffiltext{6}{University of California, Space Sciences Laboratory, 7 Gauss Way, 94720 Berkeley CA, USA}
\begin{abstract}
We study the radial evolution of intermittency of density fluctuations in the fast solar wind. The study is performed analyzing the plasma density measurements provided by Helios 2 in the inner heliosphere between $0.3$ and $0.9$ AU. The analysis is carried out by means of a complete set of diagnostic tools, including the flatness factor at different time scales to estimate intermittency, the Kolmogorov-Smirnov test to estimate the degree of intermittency, and the Fourier transform to estimate the power spectral densities of these fluctuations. Density fluctuations within fast wind are rather intermittent and their level of intermittency, together with the amplitude of intermittent events, decreases with distance from the Sun, at odds with intermittency of both magnetic field and all the other plasma parameters. Furthermore, the intermittent events are strongly correlated, exhibiting temporal clustering. This indicates that the mechanism underlying their generation departs from a time-varying Poisson process. A remarkable, qualitative similarity with the behavior of plasma density fluctuations obtained from a numerical study of the nonlinear evolution of parametric instability in the solar wind supports the idea that this mechanism has an important role in governing density fluctuations in the inner heliosphere.
\end{abstract}
\section{Introduction}
Solar wind fluctuations have been long considered as a natural realization of plasma turbulence \citep[see][and references therein]{bruno2013}. However, one of the fundamental hypotheses of the Kolmogorov's turbulence theory K41 \citep{kolmogorov1941}, i.e. the global scale invariance or self-similarity of the fluctuations, is not generally observed in space plasmas \citep[e.g.][]{burlaga1991,marsch1993,carbone1995,bruno2003}. The lack of an universal scale invariance, successively considered in Kolmogorov-Obukhov theory \citep{obukhov1962}, implies that the velocity increments measured along the flow direction $\vec{x}$ at the scale $r=|\vec{r}|$, $\delta v_{r}^{p}=\langle|\vec{V}(\vec{x}+\vec{r})-\vec{V}(\vec{x})|^{p}\rangle$, scale as $r^{s_{p}}$, i.e. $\delta v_{r}^{p}\sim r^{s_{p}}$, with the scaling exponent $s_{p}$ nonlinearly departing from the simple relation $s_{p}=p/3$, which would indicate self-similarity. The lack of self-similarity affects the shape of the Probability Density Functions (PDFs, hereafter) of the velocity increments $\delta v_{r}^{p}$. Going from larger to smaller scales, PDFs increasingly deviate from a Gaussian distribution showing fatter and fatter tails. Hence, the largest events populating the PDFs tails at the smallest scales have a larger probability to occur with respect to the Gaussian statistics. The increasing divergence from Gaussianity, as smaller and smaller scales are involved, is the typical signature of intermittency \citep{frisch1995}.

Early investigations of intermittency in the inner heliosphere \citep[e.g.][]{marsch1993,bruno2003} pointed out the different degree of intermittency of slow and fast wind, as well as of directional and compressive fluctuations, and the evolution of intermittency with the radial distance from the Sun. In particular, magnetic field fluctuations are more intermittent than velocity fluctuations and compressive fluctuations are more intermittent than directional fluctuations. Moreover, while intermittency in the slow solar wind does not show any radial dependence, both compressive and directional fluctuations in the fast solar wind become more intermittent with increasing distance.

\citet{bruno2003} interpreted these general scaling properties as a competing action between coherent (say intermittent) structures advected by the wind and stochastic (say non-intermittent) propagating Alfv$\acute{\textrm{e}}$nic fluctuations. During the wind expansion, propagating Alfv$\acute{\textrm{e}}$nic fluctuations undergo a turbulent evolution, which would reduce their Alfv$\acute{\textrm{e}}$nic character. In addition, their amplitude would be reduced faster than that of coherent structures, passively advected by the wind. In this way, the stochastic component of turbulence would progressively reduce its importance in the overall power budget of solar wind fluctuations. Thus, the net result would be a decrement of Alfv$\acute{\textrm{e}}$nicity and an increment of intermittency.

Furthermore, the question about where these coherent structures originate is still open. These structures could be generated either in the solar atmosphere and subsequently advected by the solar wind into the interplanetary medium or locally in the wind by the nonlinear evolution of turbulence \citep{bruno2001,chang2004,borovsky2008}.

Considerable efforts have been devoted to investigate the features and the behavior of magnetic field and plasma velocity fluctuations intermittency in the inner heliosphere \citep[see][for two exhaustive reviews on this topic]{tu1995,bruno2013}. However, less attention has been paid to the study of the radial evolution of the scaling properties of the interplanetary proton density fluctuations, although we have to acknowledge a careful analysis of these fluctuations at $1$ AU performed by \citet{hnat2003,hnat2005}. These authors adopted extended self similarity to reveal scaling in structure functions of density fluctuations in the solar wind and performed a comparison with scaling found in the inertial range of passive scalars in other turbulent systems. They showed that solar wind plasma density does not follow the scaling of a passive scalar. This result casted some doubts about the incompressible character of solar wind turbulence, which would follow from the continuity equation and the assumption of incompressibility $\nabla\cdot\textbf{v}=0$. The same authors pointed out that these results could also be due to the fact that direct comparison with structure functions of a passive scalar does not adequately capture the real inertial range properties of solar wind.

In addition, a distinguishing feature of density fluctuations is the flattening shown by the spectral density at high frequencies, particularly evident within fast wind but not observed within slow wind. This flattening, observed also in magnetic intensity and proton temperature spectra, assumes remarkable evidence in proton density spectra, especially for observations performed closer to the Sun \citep{marsch1990}.

The same flattening was clearly observed also in the spectrum of inward Alfv$\acute{\textrm{e}}$nic modes by \citet{grappin1990}. These authors suggested that the observed spectra could be due to non-local interaction in k-space of large-scale outward and small-scale inward Alfv$\acute{\textrm{e}}$nic modes. \citet{grappin1990} showed that a considerable amount of energy in large-scale fluctuations can be transported to small-scale fluctuations by nonlinear and non-local interactions in k-space, provided that large-scale and small-scale fluctuations have an opposite sense of propagation.

A different mechanism was suggested by \citet{tu1989} who firstly hypothesized that the spectral flattening observed in the fluctuations of inward modes and other solar wind parameters, within fast wind in the inner heliosphere, could be caused by some parametric decay processes experienced by large amplitude Alfv$\acute{\textrm{e}}$nic modes with an outward sense of propagation \citep[see also][and references therein]{matteini2012}. The same mechanism was tested by \citet{malara2001} to explain, at least qualitatively, the radial behavior of the power associated to inward and outward Alfv$\acute{\textrm{e}}$nic modes and the radial behavior of the flatness of magnetic field and velocity fluctuations \citep{primavera2003}. These authors found out that this mechanism, although capable of decreasing the alignment between magnetic and velocity fluctuations, reaches a saturation level beyond which it is no longer efficient, leaving the cross-helicity value of the fluctuations unchanged as time passes.

In this paper we investigate the radial dependence of intermittency of density fluctuations observed by Helios 2 between $0.3$ and $0.9$ AU and the effect that intermittency has on the spectral index of these fluctuations. The intermittent character of the fluctuations is firstly detected using the flatness factor, which is a good proxy of intermittency \citep{frisch1995}, and successively quantified via the Kolmogorov-Smirnov test. Then we compare our results with plasma density intermittency inferred from a compressible simulation based on the parametric decay mechanism \citep{primavera2003}, which might be related to the spectral features cited above, as suggested by \citet{tu1989}.
\section{Data selection and analysis}
The analysis is performed using $81\,s$ proton number density measurements recorded in the inner heliosphere by Helios 2 during its first solar mission in 1976. During this phase of solar cycle 21 the configuration of the polar coronal holes was remarkably stable and their meridional extensions reached very low latitudes \citep{bruno1982,villante1982}. This allowed the Helios 2 spacecraft to sample fast coronal wind in the ecliptic during three successive solar rotations \citep{bruno1985} at $0.3$, $0.7$, and $0.9$ AU, respectively. The Helios 2 in-situ measurements represent a unique dataset since these are the only measurements of the solar wind close to the Sun. This is particularly important since at $0.3$ AU, Helios 2 was observing heliospheric plasma not yet fully reprocessed by dynamical stream-stream interactions, and thus still carrying some of the pristine coronal imprints of the source region.

\begin{figure*}
	\centering
	\includegraphics[height=\hsize]{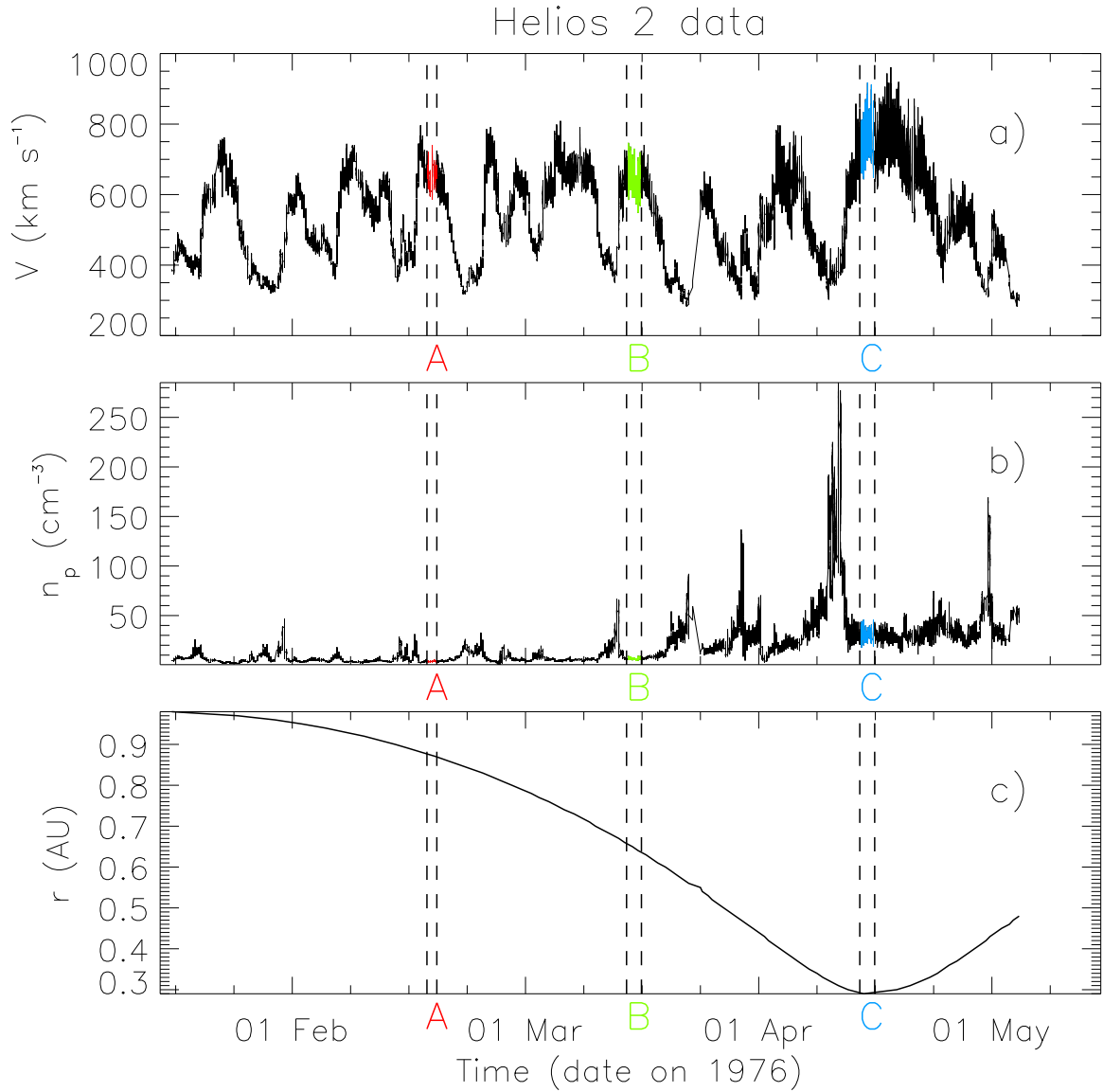}
	\caption{Time profiles of the solar wind speed \textbf{(a)} and of the plasma proton density \textbf{(b)}, and heliocentric distance of the spacecraft \textbf{(c)}, for the whole Helios 2 primary mission to the Sun; vertical dashed lines, colors red, green, and blue, and letters \lq\lq A\rq\rq, \lq\lq B\rq\rq, and \lq\lq C\rq\rq{} identify the same high-speed stream observed at three different heliocentric distances in the ecliptic plane, during three successive solar rotations.}
	\label{fig:helios_data}
\end{figure*}

The three high-speed streams, named \lq\lq A\rq\rq, \lq\lq B\rq\rq, and \lq\lq C\rq\rq, respectively, can be identified by different colors (red, green, and blue, respectively) and vertical dashed lines in Figure  \ref{fig:helios_data}, which displays, from top to bottom, the time profiles of the solar wind speed \textbf{(Figure  \ref{fig:helios_data}a)} and of the plasma proton density \textbf{(Figure  \ref{fig:helios_data}b)}, and the spacecraft heliocentric distance \textbf{(Figure  \ref{fig:helios_data}c)}, for the whole Helios 2 primary mission to the Sun.

The beginning and the end of each selected temporal interval, lasting about 2 days each, its duration $\Delta t$, the average heliocentric distance $\langle r\rangle$, wind speed $\langle V\rangle$, and proton density $\langle n_{p}\rangle$ are listed in Table \ref{table:helios_data}.

\begin{table*}
	\begin{center}
		\begin{tabular}{ccccc}
			\tableline\tableline
			Time interval & $\Delta t$ & $\langle r\rangle$ & $\langle V\rangle$ & $\langle n_{p}\rangle$ \\
			YYYY MM DD, hh:mm & [h] & [AU] & [$km\,s^{-1}$] & [$cm^{-3}$] \\
			\tableline
			1976 02 18, 19:40 - 1976 02 20, 02:22 & 31 & 0.88 & 644 & 3.7 \\
			1976 03 15, 12:01 - 1976 03 17, 10:04 & 46 & 0.65 & 630 & 6.4 \\
			1976 04 14, 12:01 - 1976 04 16, 10:04 & 46 & 0.29 & 735 & 30.5 \\
			\tableline\tableline
		\end{tabular}
		\caption{Beginning, end, and duration of the selected temporal intervals, average heliocentric distance, wind speed, and proton density, referring to the same high-speed stream observed at three different heliocentric distances in the ecliptic plane, during three successive solar rotations.}
		\label{table:helios_data}
	\end{center}
\end{table*}

These three fast wind streams are notorious for being dominated by Alfv$\acute{\textrm{e}}$nic fluctuations and have been extensively studied since they offer the unique opportunity to observe the radial evolution of MHD turbulence within the inner heliosphere \citep[see][and references therein]{bruno2013}. As anticipated in the previous section, these streams have been considered for studying the radial evolution of the intermittency of proton number density fluctuations.

The PDFs of a fluctuating field affected by intermittency become more and more peaked when analyzing smaller and smaller scales \citep{frisch1995}. Since the peakedness of a distribution is measured by its flatness factor $\mathcal{F}$ we examined the behavior of this parameter at different scales to unravel the presence of intermittency in our time series. It can be assumed that fluctuations are more intermittent as either $\mathcal{F}$ increases faster with changing scales or starts to increase at larger scales.

The flatness factor $\mathcal{F}$ at the scale $\tau$ is defined as

\begin{equation}
	\mathcal{F}(\tau)=\frac{\langle\delta\rho_{\tau}^{4}\rangle}{\langle\delta\rho_{\tau}^{2}\rangle^{2}},
	\label{eq:flatness_factor}
\end{equation}

where $\delta\rho_{\tau}=\rho(t+\tau)-\rho(t)$ are the density fluctuations at the scale $\tau$, and the brackets indicate an averaging done over the time interval considered.

\begin{figure*}
	\centering
	\includegraphics[height=\hsize,angle=90]{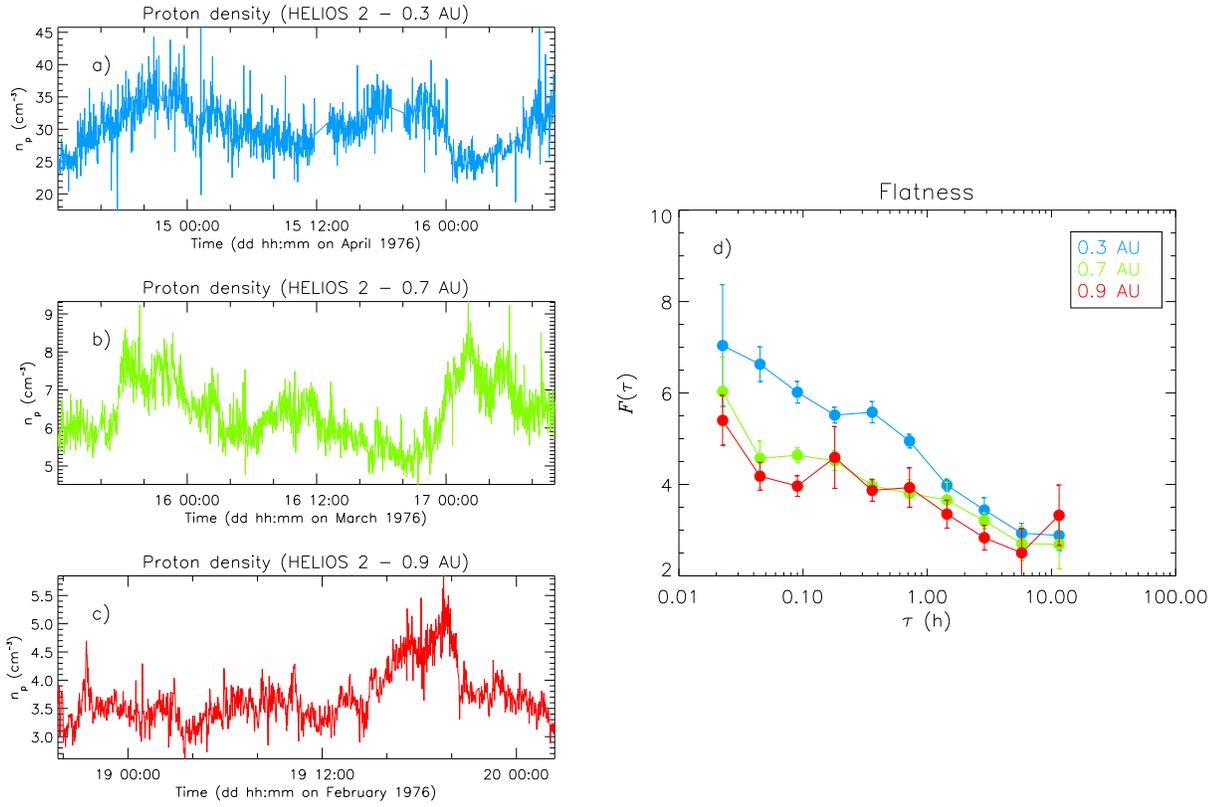}
	\caption{Time profiles of the proton density measured by Helios 2 at $0.3$ \textbf{(a)}, $0.7$ \textbf{(b)}, and $0.9$ \textbf{(c)} AU, within the same corotating high-speed stream; data gaps are removed by linear interpolation. \textbf{(d)} Flatness factor $\mathcal{F}$ as a function of time scale $\tau$, relative to fluctuations of the plasma density observed in fast wind, at $0.3$, $0.7$, and $0.9$ AU (blue, green, and red curves, respectively).}
	\label{fig:flatness_factor}
\end{figure*}

The values of the flatness factor $\mathcal{F}$ of the density fluctuations observed in the fast solar wind at $0.3$, $0.7$, and $0.9$ AU, as computed by using Eq. \ref{eq:flatness_factor}, are shown in \textbf{Figure  \ref{fig:flatness_factor}d} as a function of time scale $\tau$; the time profiles of the proton density measured by Helios 2 at the three different heliocentric distances \textbf{are displayed in Figures  \ref{fig:flatness_factor}a,b,c}.

The flatness factor $\mathcal{F}$ for each data sample grows moving from larger to smaller scales, unraveling the intermittent character of the fluctuations. However, since at $0.3$ AU the flatness starts to increase at larger scales and grows more rapidly, reaching higher values at small scales (blue curve in \textbf{Figure  \ref{fig:flatness_factor}d}), density fluctuations at $0.3$ AU can be considered more intermittent than those at $0.7$ and $0.9$ AU. Moreover, the density fluctuations at $0.7$ and $0.9$ AU exhibit approximately the same level of intermittency, since the corresponding $\mathcal{F}$ curves overlap to each other along the whole range of scales, within the associated uncertainties (green and red curves, respectively, in \textbf{Figure  \ref{fig:flatness_factor}d}). Thus, at least in the transition from $0.3$ to $0.7$ AU, there is a clear evidence for a radial depletion of intermittency. This result, confirmed by Helios 1 observations in 1975 and not shown in this paper, are at odds with the intermittent behavior of all the other field and plasma parameters, which become more intermittent when the heliocentric distance increases in the inner heliosphere \citep{tu1995,bruno2013}.

\begin{figure*}
	\centering
	\includegraphics[height=\hsize]{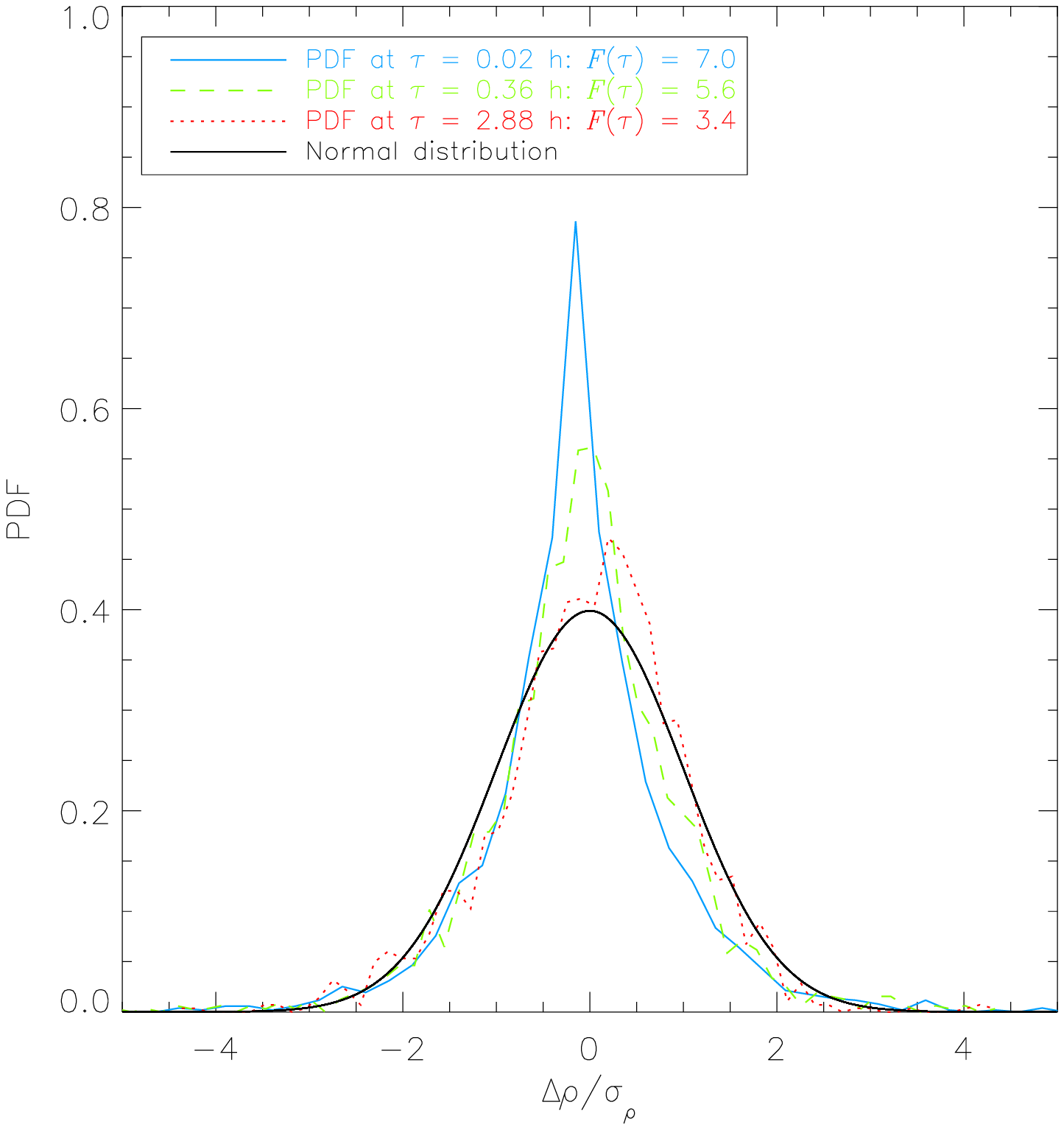}
	\caption{Probability distribution functions of proton density increments normalized to the standard deviation $\Delta\rho/\sigma_{\rho}$, obtained at $0.3$ AU, at the scales $\tau=0.02\,h$ (blue continuous curve), $\tau=0.36\,h$ (green dashed curve), and $\tau=2.88\,h$ (red dotted curve); the corresponding flatness factors are reported in the legend; a Gaussian distribution function with average $\mu=0$ and standard deviation $\sigma=1$ is displayed for reference (black curve).}
	\label{fig:pdf}
\end{figure*}

In order to show how the intermittency affects the shape of the distributions of density fluctuations and, as a consequence the corresponding flatness, the PDFs of proton density increments $\Delta\rho$ normalized to the standard deviation $\sigma_{\rho}$, obtained at $0.3$ AU and at three different scales, are displayed in Figure  \ref{fig:pdf} where they are also compared with a normal distribution.

As smaller and smaller scales are involved, going from the larger scale $\tau=2.88\,h$ (red dotted curve in Figure  \ref{fig:pdf}) to the intermediate scale $\tau=0.36\,h$ (green dashed curve in Figure  \ref{fig:pdf}) and, finally to the smallest one $\tau=0.02\,h$ (blue continuous curve in Figure  \ref{fig:pdf}), the PDFs progressively deviate from a Gaussian distribution (black curve in Figure  \ref{fig:pdf}), becoming more and more peaked. This is reflected into an increase of the flatness factor from $\mathcal{F}(\tau=2.88\,h)=3.4$ (close to the value of a normal distribution) to $\mathcal{F}(\tau=2.88\,h)=7.0$.

The study of $\mathcal{F}$ gives only a partial description of the intermittency behavior and might also be not particularly sensitive to distinguishing small differences in the degree of intermittency, especially for data sample collected at heliocentric distances close to each other, as for the proton density measured at $0.7$ and $0.9$ AU, whose intermittency might not have had sufficient time to evolve significantly. In order to quantify the degree of intermittency present in our selected intervals we performed the Kolmogorov-Smirnov (K-S, hereafter) test which allows us to investigate the statistics of persistence between intermittent events and to characterize their temporal clustering.

However, before proceeding with the K-S test we adopted the Local Intermittent Measure (LIM, hereafter) technique to localize intermittent events in scale and time.
\subsection{The Local Intermittency Measure}
This method, firstly introduced by \citet{farge1992}, relays on a wavelet decomposition of the signal. The wavelet analysis is indeed a powerful tool for capturing peaks, discontinuities, and sharp variations in the input data. The LIM at the scale of interest $\tau$ is defined as \citep{farge1992}

\begin{equation}
	\mathrm{LIM}(\tau,t)=\frac{\omega(\tau,t)^{2}}{\langle|\omega(\tau,t)|^{2}\rangle}_{t},
	\label{eq:lim}
\end{equation}

where $\omega(\tau,t)$ are the coefficients of the wavelet transform. The expression in Eq. \ref{eq:lim} hence represents the energy content of the density fluctuations at the scale $\tau$ and time $t$ normalized to the average value of the energy at the same scale over the whole data sample. This average value corresponds to the estimate that one would obtain at that scale from a Fourier spectrum. Then, $\mathrm{LIM}(\tau,t)>1$ indicates that a given scale at a given time contributes more than the average over the entire data sample to the Fourier spectrum at scale $\tau$. However, a more direct indicator of intermittency is represented by $\mathrm{LIM}^{2}$ averaged on the whole time interval, which defines the flatness $\mathcal{F}_{w}$ of the wavelet coefficients \citep{meneveau1991} for each scale $\tau$.

This method, used to identify intermittent events, is based upon a recursive procedure \citep{bianchini1999}. For each scale $\tau$ we compute the value of $\mathcal{F}_{w}$. If this value is $>3$ we eliminate $3\%$ of the outliers from the distribution of the wavelet coefficients at that scale, reconstruct the signal and recompute $\mathcal{F}_{w}$. If $\mathcal{F}_{w}$ is still $>3$ we eliminate again $3\%$ of the outliers and proceed recursively, as in the previous cycle, until $\mathcal{F}_{w}$ does not change any longer. Collecting all the intermittent events, i.e. the outliers, allows us to build their time series and investigate their Waiting Time Distributions (WTDs, hereafter) via the K-S test described below.
\subsection{The Kolmogorov-Smirnov test}
This method allows us to quantify the degree of correlation of the density fluctuations or, in other words, their level of intermittency. The K-S test is indeed a nonparametric and distribution-free test used to compare a data sample with a reference probability distribution, by quantifying the maximum distance $D$ between the empirical and the reference distribution function. In the analysis of the time distribution of the intermittent events, we tested whether or not the sequence of residuals was consistent with a time-varying Poisson process \citep{bi1989,carbone2004,carbone2006}.

Starting from the sequence of time intervals between two successive residuals $\delta t$ (waiting time sequence) as a function of the intermittent event occurrence time $t$, the stochastic variable $h$ can be defined as:

\begin{equation}
	h_{i}(\delta t_{i},\delta\tau_{i})=\frac{2\delta t_{i}}{2\delta t_{i}+\delta\tau_{i}},
	\label{eq:h}
\end{equation}

where $\delta t_{i}$ and $\delta\tau_{i}$ are the waiting times between an intermittent event at $t_{i}$ and the two (either following or preceding) nearest outliers:

\begin{equation}
	\delta t_{i}=\min\left\{t_{i+1}-t_{i},t_{i}-t_{i-1}\right\}
	\label{eq:delta_t}
\end{equation}

\begin{equation}
	\delta\tau_{i}=\left\{\begin{array}{ll}
	t_{i-1}-t_{i-2} & \mathrm{if}\,\delta t_{i}=t_{i}-t_{i-1}\\
	t_{i+2}-t_{i+1} & \mathrm{if}\,\delta t_{i}=t_{i+1}-t_{i}
	\end{array}\right.
	\label{eq:delta_tau}
\end{equation}

The stochastic variable $h$ hence simply represents the suitably normalized time between intermittent events. Under the null hypothesis that the data sample is drawn from the Poisson reference distribution, that is in the hypothesis that $\delta t_{i}$ and $\delta\tau_{i}$ are independently distributed with exponential probability densities given by $p(\delta t_{i})=2\nu_{i}\exp(-2\nu_{i}\delta t_{i})$ and $p(\delta\tau_{i})=\nu_{i}\exp(-\nu_{i}\delta\tau_{i})$, where $\nu_{i}$ is the instant event rate, it can be easily shown that the Cumulative Distribution Function (CDF, hereafter) of $h$, $P(h<H)$, is simply $P(h<H)=H$, where $P(h)$ is the probability distribution function of $h$. Hence, if the Poisson hypothesis holds, the stochastic variable $h$ is uniformly distributed in $[0:1]$. The maximum deviation $D$ of the empirical CDF of the outliers from the reference relation $P(h<H)=H$, quantifies the degree of departure of the intermittent events from the Poissonian, i.e. non-intermittent, statistics.
\subsection{The Waiting Time Distribution of intermittent events}
\begin{figure*}
	\centering
	\includegraphics[height=\hsize,angle=90]{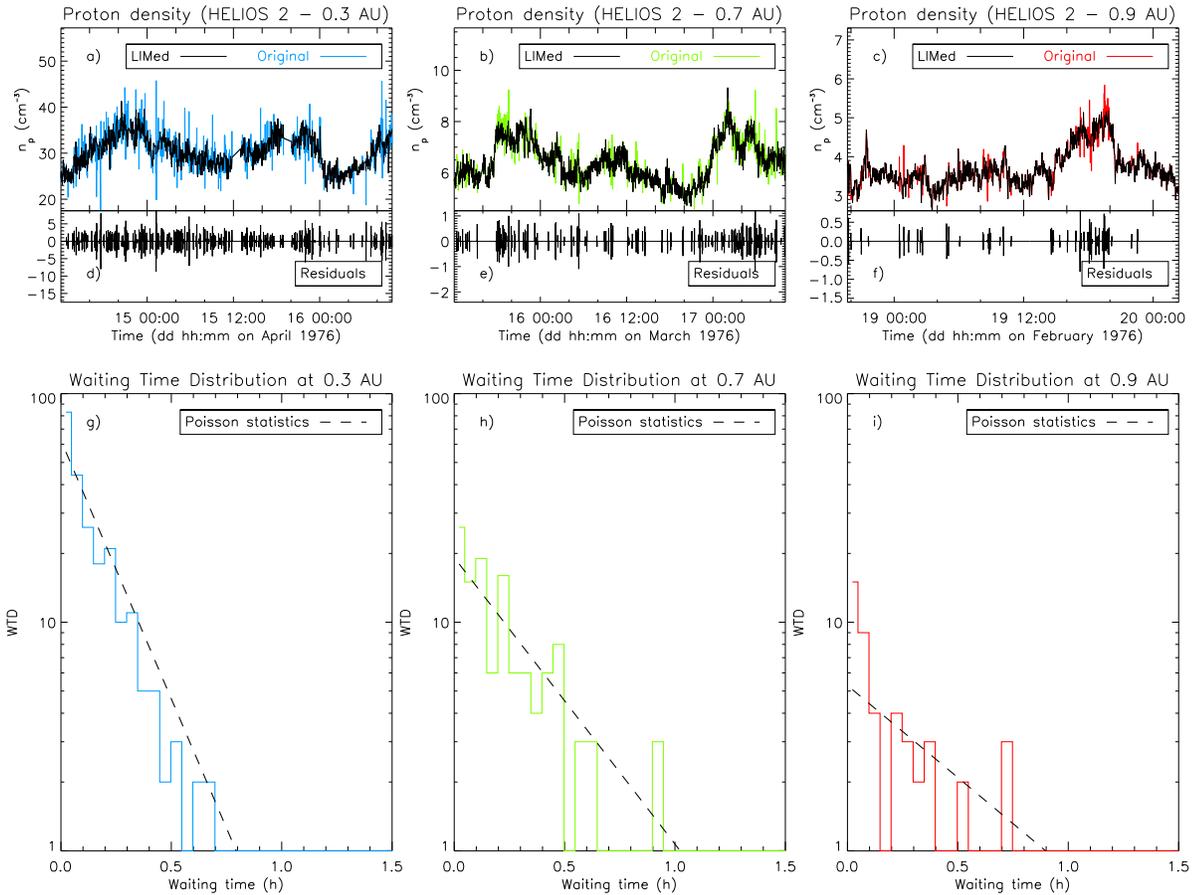}
	\caption{Time profiles of the original data collected by Helios 2 (data gaps were linearly interpolated) \textbf{at $0.3$ (a), $0.7$ (b), and $0.9$ AU (c)}, with superimposed the trace of the LIMed signal (colored and black curves, respectively). Intermittent component obtained from the difference between the original and the LIMed data, \textbf{at $0.3$ (d), $0.7$ (e), and $0.9$ AU (f)}. Waiting time distributions between two consecutive intermittent events as a function of time scale $\tau$\textbf{, as identified at $0.3$ (g), $0.7$ (h), and $0.9$ AU (i)}; the theoretical distribution function expected under a Poisson statistics is also shown (dashed lines). Colors are the same adopted for Figures  \ref{fig:helios_data} and \ref{fig:flatness_factor}.}
	\label{fig:lim}
\end{figure*}

\begin{figure*}
	\centering
	\includegraphics[height=\hsize,angle=90]{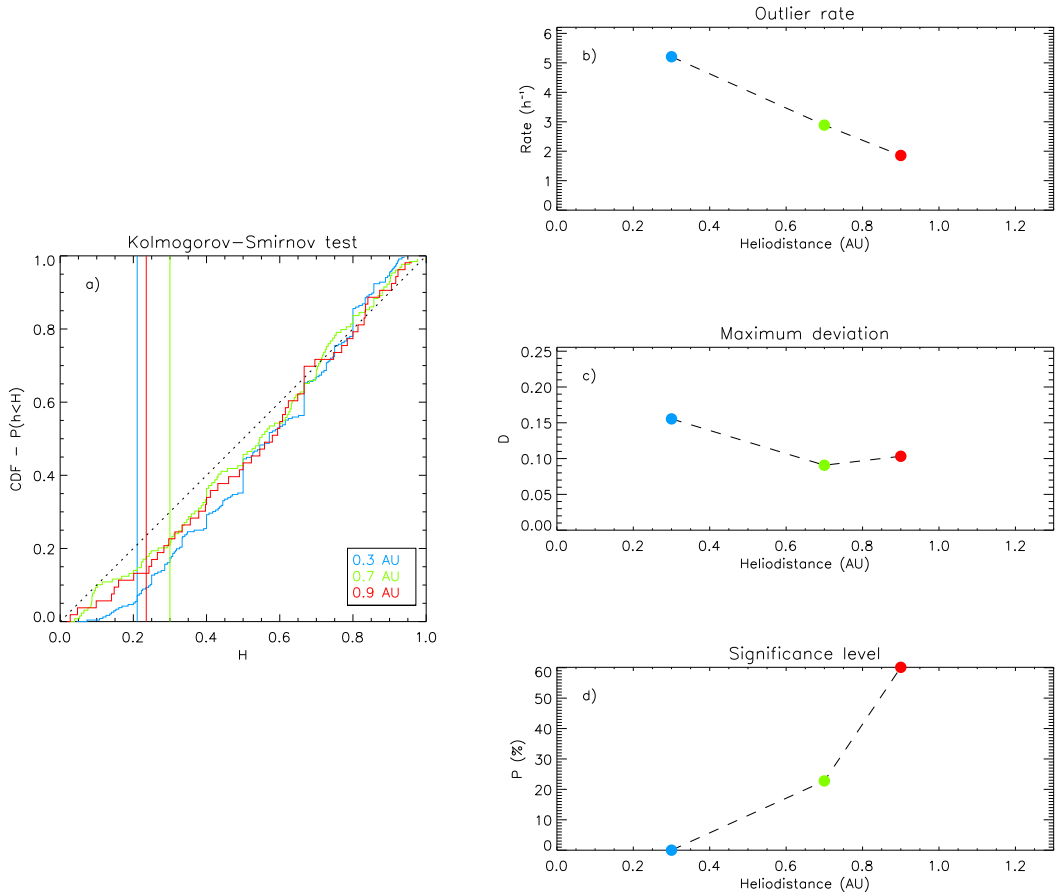}
	\caption{\textbf{(a)} Cumulative distribution functions $P(h<H)$ of the normalized time $h$ between intermittent events identified at $0.3$, $0.7$, and $0.9$ AU (blue, green, and red curves, respectively); the dotted line represents the theoretical probability expected under a Poisson statistics; the vertical lines indicate where the CDFs mostly deviate from it (the three different colors refer to the different heliocentric distances as reported in the legend). Occurrence rate $\nu$ of the intermittent events \textbf{(b)}, maximum deviation $D$, deduced \textbf{(a)}, between the empirical and the reference (say Poisson) CDFs \textbf{(c)}, and significance level $P$ of the K-S test \textbf{(d)}, as functions of the heliocentric distance; colors are the same used for \textbf{(a)}.}
	\label{fig:k-s_test}
\end{figure*}

The intermittent events identified by the LIM technique are shown in \textbf{Figures \ref{fig:lim}a,b,c}, while \textbf{Figures \ref{fig:lim}ghi} and Figure \ref{fig:k-s_test} show the WTDs and the results of the K-S test, respectively.

The colored time profiles in \textbf{Figures  \ref{fig:lim}a,b,c} show the original proton density data recorded by Helios 2 within fast wind at $0.3$, $0.7$, and $0.9$ AU. The superimposed black curves represent the LIMed data, i.e. the same time series after the intermittent events have been filtered out via the LIM technique: the residuals, shown below \textbf{in Figures  \ref{fig:lim}def} as the difference between the two curves, are mainly made of most of the peaks of the original data.

The WTDs (\textbf{Figures  \ref{fig:lim}g,h,i}) between two subsequent outliers have been tested for consistency with Poisson statistics. Under the Poisson hypothesis, the delay time $\Delta t$ between two consecutive intermittent events should be distributed according to the following function:

\begin{equation}
	n(\Delta t)=N_{out}\nu_{out}\exp(-\Delta t\cdot\nu_{out}),
	\label{eq:poisson_wtd}
\end{equation}

where $N_{out}$ is the number of outliers, say intermittent events, identified via LIM in the original data and $\nu_{out}$ is the occurrence rate of the outliers computed as the total number of outliers divided by the time duration of the interval. In particular, $\nu_{out}$ resulted to be $5.2\,h^{-1}$, $2.9\,h^{-1}$, and $1.9\,h^{-1}$ at $0.3$, $0.7$, and $0.9$ AU, respectively, showing a clear decreasing trend with increasing the distance from the Sun \textbf{(Figure  \ref{fig:k-s_test}b)}.

It is readily seen that the WTD of the intermittent events identified in the density fluctuations at $0.3$ AU is systematically smaller than the theoretical distribution function expected under a Poisson statistics, indicating that the events are not stochastically distributed and that correlated clusters are present in the time sequence of the outliers. On the other hand, due to the lower statistics of the intermittent events at $0.7$ and $0.9$ AU, it is not clear whether or not the WTDs of the residuals are consistent with a time-varying Poisson process. As we will see in the following, the K-S test provides a quantitative description of clustering properties of the intermittent events allowing us to quantify the degree of intermittency.

The CDF $P(h<H)$ of the normalized time $h$ between the intermittent events identified at $0.3$, $0.7$, and $0.9$ AU are shown in \textbf{Figure  \ref{fig:k-s_test}a} where they are compared with the theoretical Poisson probability function (dotted lines). All of these cumulative distribution functions $P(h<H)$ strongly depart from the Poisson theoretical distribution $P(h<H)=H$, being below it for $H<1/2$. These observational evidences represent unambiguous indications of temporal clustering of the intermittent events. The clusters in the outlier sequences are evident also looking at the corresponding time profiles (\textbf{Figures  \ref{fig:lim}a,b,c}). The maximum distance $D$ between the empirical and the reference distributions is shown, as a function of heliocentric distance, in \textbf{Figure  \ref{fig:k-s_test}c}. The low values of the significance level of the K-S test (displayed as a function of the heliocentric distance in \textbf{Figure  \ref{fig:k-s_test}d}) highlight and quantify the presence of correlations among the intermittent events, which lead to their clustering, and suggest that these correlations are stronger closer to the Sun. In other words, the departure from the Poisson statistics, i.e. the probability that the intermittent events are not dominated by a Poisson process, becomes stronger and stronger with decreasing the heliocentric distance from $0.9$ to $0.3$ AU with the lowest value ($P\sim0\%$) registered at $0.3$ AU.

\subsection{The spectral analysis}
\begin{figure*}
	\centering
	\includegraphics[height=\hsize]{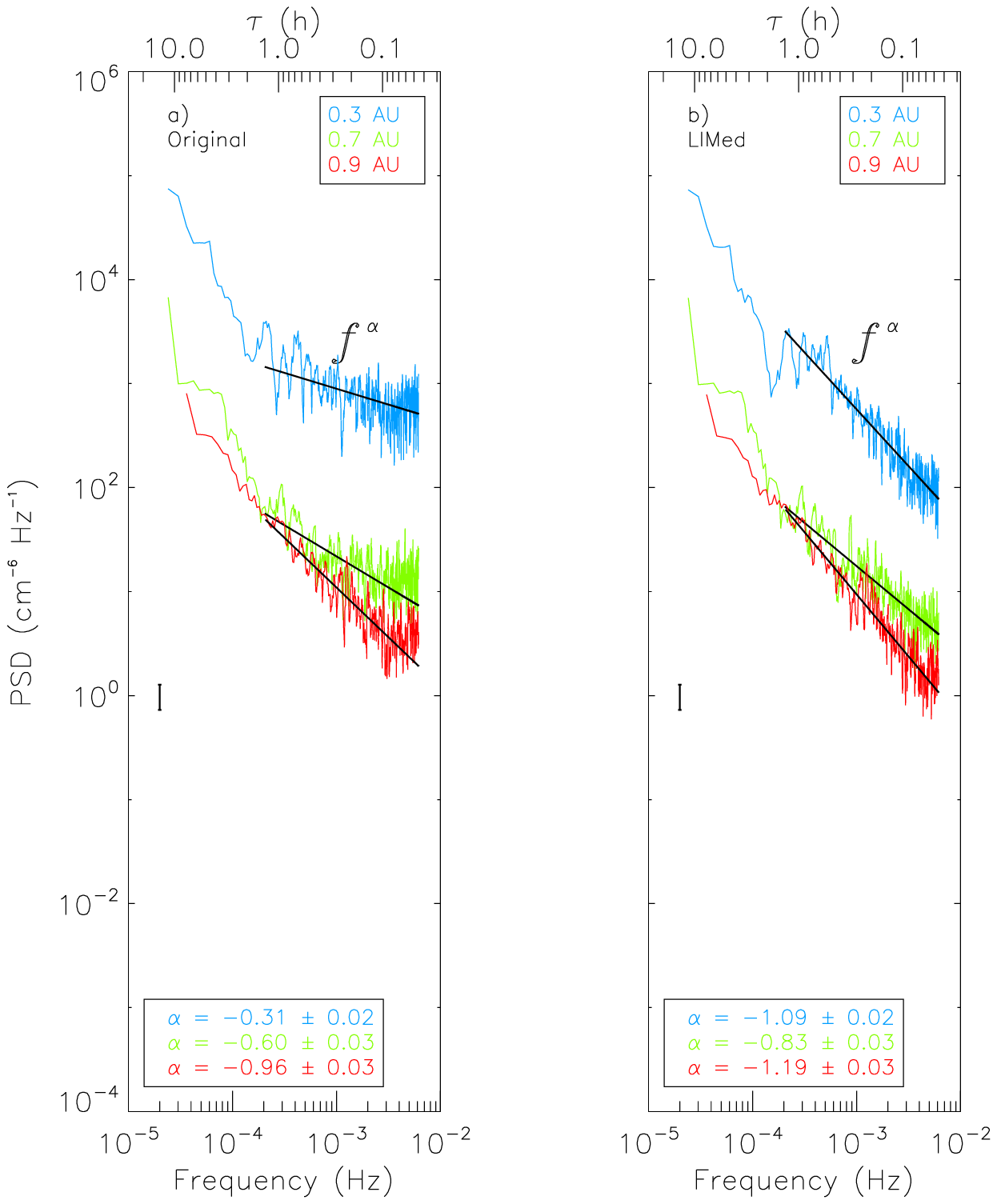}
	\caption{Power spectral density vs frequency relative to fluctuations of the original \textbf{(a) and LIMed (b)} density time series observed in the fast solar wind at $0.3$, $0.7$, and $0.9$ AU; thick solid lines show the $f^{\alpha}$ slopes as inferred by the fitting functions; the spectral slopes $\alpha$ relative to the three different heliocentric distances are reported in the legend; the error bars indicate a confidence level of 95\%. Colors are the same used for Figures  \ref{fig:helios_data}, \ref{fig:flatness_factor}, and \ref{fig:lim}.}
	\label{fig:fft}
\end{figure*}

We evaluated the effect of intermittency of the density fluctuations on the Power Spectral Density (PSD, hereafter) of these fluctuations. \textbf{Figures \ref{fig:fft}a,b} show the PSD of the fluctuations of the original and LIMed proton density time series acquired in the fast solar wind at $0.3$, $0.7$, and $0.9$ AU.

The spectral analysis of the original Helios 2 data at the three different heliocentric distances (\textbf{Figure  \ref{fig:fft}a}) reveals the existence of large-scale fluctuations of the proton density in the fast solar wind, over time scales ranging from $\sim1$ to $\sim10\,h$, where they exhibit approximately the same low-frequency spectral scaling close to the $1/f^{2}$ Brownian noise. The occurrence of $1/f^{2}$ noise \citep[widely observed in the interplanetary medium,][]{sari1969,coles1978,burlaga1987,roberts1987} is indicative of the presence of uncorrelated discontinuities, likely generated by stochastic processes close to a Brownian noise. Structures of this kind could arise from shocks, compressive modes or inhomogeneities embedded in the solar wind, such as current sheets like tangential discontinuities separating adjacent regions characterized by different total pressure and density \citep{bruno2013}. The characteristics of density fluctuations within fast wind at low frequencies are maintained in the transition from $0.3$ out to $0.9$ AU. In contrast, a quick dynamical evolution of the density fluctuations is observed at higher frequencies during the wind expansion. Indeed, for frequencies larger than $\sim3\times10^{-4}$ Hz, i.e. for time scales smaller than $\sim1\,h$, the spectra of the original data flattens out to remarkable low values of the spectral slope. However, this part of the spectrum becomes steeper and steeper moving away from the Sun. The spectral slopes, as inferred by a $f^{\alpha}$ fit in this high frequency range (by applying the Levenberg-Marquardt least-squares minimization method with a confidence level of 95\%) are $\alpha=-0.31\pm0.02$, $\alpha=-0.60\pm0.03$, and $\alpha=-0.96\pm0.03$ at $0.3$, $0.7$, and $0.9$ AU, respectively.

The flattening of these spectra starts roughly at the same scales where the flatness factor $\mathcal{F}$ starts to increase and both the spectral slope and the flatness factor $\mathcal{F}$ depend on the radial distance from the Sun. Thus, we computed the PSDs of the LIMed time series, which are shown in \textbf{Figure  \ref{fig:fft}b}. The comparison with the spectra shown in \textbf{Figure  \ref{fig:fft}a} unravels that most of the flattening was due to the intermittent events and that the radial dependence is much reduced. The spectral slopes within the high frequency range are all roughly around $-1$, which indicate that the time series is statistically uncorrelated, in agreement with the expected results from the LIM technique operated on the original data. In addition, the largest correction was experienced by the time series that was the most intermittent, i.e. the one at $0.3$ AU.

These results suggest that intermittent events might be the result of some mechanism whose efficiency decreases with increasing the heliocentric distance. \citet{tu1989} firstly suggested that the flattening observed in the high frequency termination of proton number density PSDs observed by Helios were probably due to parametric decay instability experienced by large amplitude Alfv$\acute{\textrm{e}}$n waves, particularly active within high speed streams \citep{tu1995,bavassano2000,primavera2003,bruno2013}.


Inward propagating Alfv$\acute{\textrm{e}}$nic modes, necessary to start the nonlinear interactions with the outward propagating counterpart, would be generated locally by the parametric decay of outward modes, which are unstable to density perturbations and would indeed decay into a backscattered Alfv$\acute{\textrm{e}}$n waves with lower amplitude and frequency (which would produce a decrease of the initial Alfv$\acute{\textrm{e}}$nic correlation of the fluctuations) and into an acoustic wave propagating in the same direction of the mother wave. This last compressive component would enhance the density spectra at high frequency, as observed in \textbf{Figure  \ref{fig:fft}a}. However, the amplitude of the outward propagating Alfv$\acute{\textrm{e}}$n waves, which initially dominate the fast solar wind, becomes weaker and weaker with increasing distance from the Sun. Hence, the parametric instability is expected to become less and less efficient during the nonlinear evolution from $0.3$ out to $0.9$ AU, until it reaches a saturation state. It turns out that both the number of intermittent events generated per unit time via parametric instability and, in turn, the level of intermittency of the density fluctuations is expected to reduce throughout the inner heliosphere, as observed in Figures \ref{fig:lim}, \ref{fig:k-s_test}, and \ref{fig:fft}. Hence, the parametric instability might account for the observed radial evolution of intermittency found in density fluctuations within the fast solar wind.
\subsection{Numerical Simulations}
As a further check we can compare, at least qualitatively, the intermittent character of the observed density fluctuations with intermittency found in parametric decay simulations performed by \citet{primavera2003}, but not reported in their original paper.

These authors used a pseudo-spectral, one dimensional, compressible MHD code to simulate the evolution of an Alfv$\acute{\textrm{e}}$n wave with an extended spectrum in a uniform background magnetic field $\textbf{B}_{0}$. The boundary conditions are periodicity on the spatial domain $0\le x<2\pi$, where $x$ is a dimensionless spatial coordinate, normalized to a length $L$, which could represent the correlation length in turbulence. In the initial condition the perturbations transverse to $\textbf{B}_{0}$ of the magnetic field, $\delta\textbf{B}_{\perp}$, and velocity, $\delta\textbf{v}_{\perp}$, are correlated as in a forward-propagating Alfv$\acute{\textrm{e}}$n wave: $\delta\textbf{v}_{\perp}=-(c_{A0}/B_{0})\delta\textbf{B}_{\perp}$, where $c_{A0}$ is the uniform background Alfv$\acute{\textrm{e}}$n velocity. Moreover, $\delta\textbf{v}_{\perp}$ and $\delta\textbf{B}_{\perp}$ have a uniform modulus, while they change direction along the spatial domain:

\begin{equation}
	\delta\textbf{B}_{\perp}(x,t=0)=B_{1}\lbrace\cos\lbrack\phi(x)\rbrack\textbf{e}_{y}+\sin\lbrack\phi(x)\rbrack\textbf{e}_{z}\rbrace,
	\label{eq:delta_b}
\end{equation}

where $B_{1}=B_{0}/2$ is the wave amplitude, $\textbf{e}_{y}$ and $\textbf{e}_{z}$ are two unit vectors perpendicular to the propagation direction $x$, and the phase $\phi(x)$ is a periodic function with a given spectrum; as a consequence, both $\textbf{v}_{\perp}$ and $\textbf{B}_{\perp}$ have a spectrum of wavelengths covering several Fourier harmonics. This perturbation is an exact solution of the ideal MHD equations, i.e. when neglecting dissipation it propagates undistorted along $\textbf{B}_{0}$ at the constant Alfv$\acute{\textrm{e}}$n speed $c_{A0}$. However, this solution is unstable with respect to the generation of both a backward-propagating Alfv$\acute{\textrm{e}}$nic perturbation and a compressive perturbation \citep{malara1996}. This instability represents the extension of the single-wavelength parametric instability \citep{goldstein1978} to a multi-wavelength configuration. In the simulation the parametric instability is triggered by adding a low amplitude noise to the initial Alfv$\acute{\textrm{e}}$nic perturbation, and the evolution is followed throughout the linear stage of the instability and after the saturation. Values of plasma $\beta\sim1$ are used, which are close to the observed values \citep{tu1995}. The results show that the initial alignment between $\delta\textbf{B}_{\perp}$ and $\delta\textbf{v}_{\perp}$ is progressively destroyed till energy of the forward and backward propagating Alfv$\acute{\textrm{e}}$n waves become comparable, while the amplitude of density perturbation grows to a level $\delta\rho\sim0.03\rho_{0}$, where $\rho_{0}$ is the background density. This corresponds to the instability saturation. At that time velocity, magnetic field and density have developed extended power-law spectra. At subsequent times, velocity and magnetic field fluctuations became more intermittent with increasing time, showing qualitative analogy with observations by \citet{bruno2003} recorded at different heliocentric distances.

\begin{figure*}
	\centering
	\includegraphics[width=\hsize]{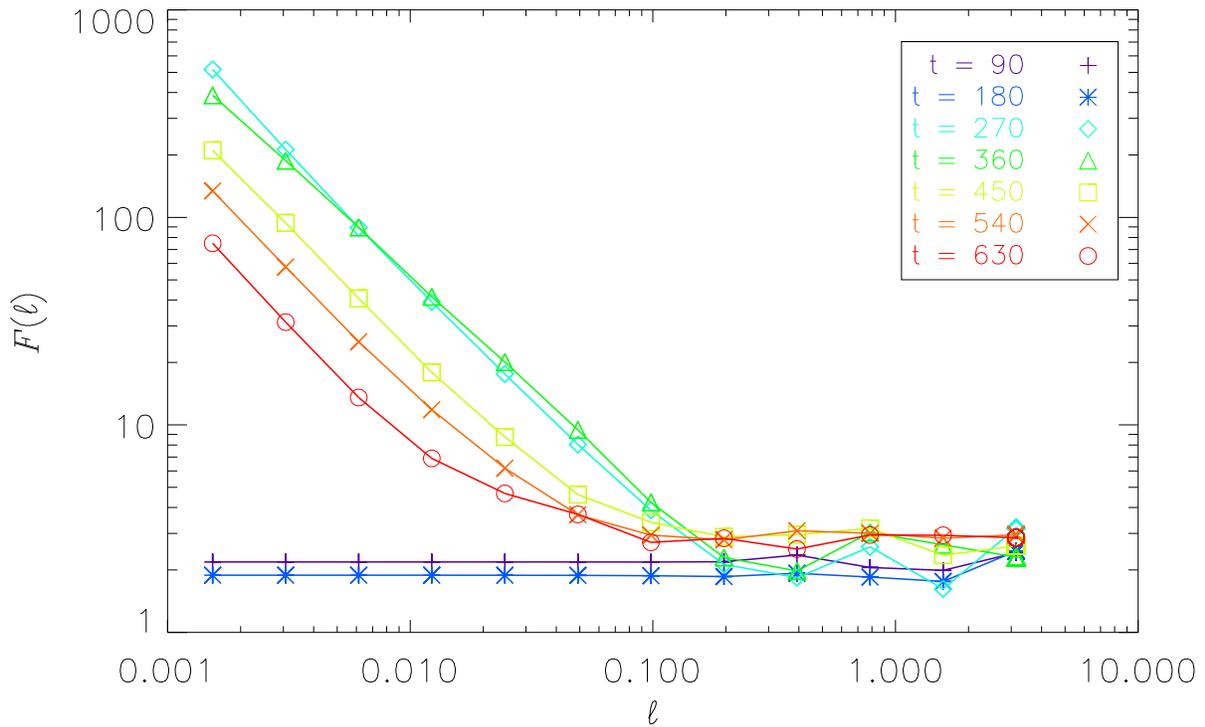}
	\caption{Flatness factor $\mathcal{F(\ell)}$ of density fluctuations, obtained from numerical simulation, as a function of the spatial scale $\ell$. Different curves correspond to different times during the simulation, as reported in the legend. The unit time is the Alfv$\acute{\textrm{e}}$n time, namely the time it takes an Alfv$\acute{\textrm{e}}$n fluctuation to cross the simulation domain.}
	\label{fig:flatness_factor_mhd}
\end{figure*}

In the present paper we examined the behavior of the flatness factor $\mathcal{F}$ of density fluctuations obtained from the above-described simulation, which has been carried out for times longer than those considered in \citet{primavera2003}. Of course, in this case $\mathcal{F}$ is to be considered a function of the spatial scale $\ell$ instead of the time interval $\tau$. In Figure  \ref{fig:flatness_factor_mhd} $\mathcal{F}$ is plotted as a function of $\ell$.

The seven different curves refer to as many successive times (the unit time is the Alfv$\acute{\textrm{e}}$n time, namely the time it takes an Alfv$\acute{\textrm{e}}$n fluctuation to cross the simulation domain) during the simulation. At the instability saturation $\mathcal{F}$ has a value close to $2$ throughout the whole range of scales. At that time fluctuations have still a coherent character due to the wave interaction that has governed the parametric instability, so that the turbulence is not yet fully developed. At a subsequent time ($t\sim270$) in the small scale range $\ell\le0.1$ the profile of $\mathcal{F}$ strongly increases with decreasing the scale $\ell$, while $\mathcal{F}$ fluctuates around the value $3$ at large scales, $\ell\ge0.1$. After that time, $\mathcal{F}$ starts to decrease in the small scale range, while keeping the same value at large scales. This behavior is qualitatively similar to our observations if we base this comparison on the equivalence between time in the simulation and radial distance in the in-situ observations.
\section{Conclusions}
In this article we analyzed the proton density measurements collected by Helios 2 within the same corotating high-speed stream observed at $0.3$, $0.7$, and $0.9$ AU, aiming to investigate the radial evolution of intermittency in fast solar wind turbulence.


The evaluation of the flatness factor, the wavelet-based identification of intermittent events and the quantitative description of their temporal clustering (indicating presence of strong correlations) have shown evidence that density intermittency decreases with heliocentric distance. This behavior is opposite to the typical evolution of other solar wind plasma and magnetic field parameters, for which intermittency increases with radial distance. In the range of scales where intermittency is stronger, density spectra show a remarkable flattening, which decreases moving away from the Sun. The removal of the intermittent events from the time series dramatically reduces such spectral flattening, canceling de facto its radial dependence.

The decrease of density intermittency with heliocentric distance and the spectral properties are in agreement with the results of pseudo-spectral, one dimensional, MHD numerical simulation of the propagation of an extended turbulent spectrum of Alfv$\acute{\textrm{e}}$n waves in a uniform background magnetic field. The parametric decay of large amplitude Alfv$\acute{\textrm{e}}$n waves could therefore play a relevant role in the radial evolution of solar wind turbulence, representing the candidate mechanism able to explain the observations described in this study.


Parametric instability drives an outward large amplitude Alfv$\acute{\textrm{e}}$n wave to decay into an inward Alfv$\acute{\textrm{e}}$n mode at smaller wave vector and a compressive wave propagating in the same direction of the pump wave. The observed spectral flattening \citep{tu1989} and the subsequent spectral steepening with radial distance is compatible with the progressive saturation of parametric decay mechanism during the solar wind expansion.
\begin{acknowledgements}
The research leading to these results has received funding from the European Community's Seventh Framework Programme (FP7/2007-2013) under grant agreements N. 313038/STORM and N. 269297/TURBOPLASMAS. This work was also supported by the Italian Space Agency (ASI) grants (I/013/12/0). The financial contribution from the ASI agreement N. 11013112/0 is acknowledged. The authors thank the anonymous Referee for his/her valuable comments and suggestions.
\end{acknowledgements}

\end{document}